\documentclass[twocolumn,secnumarabic,amssymb, nobibnotes, aps]{revtex4}
\usepackage{hyperref}
\usepackage{graphicx}
\usepackage{amsmath,amssymb,float}
\usepackage{epstopdf}

\begin{document}

\title{Electronic Trap States in Methanofullerenes}

\author{Julia Schafferhans $^1$}
\email[Electronic mail: ]{julia.schafferhans@physik.uni-wuerzburg.de}
\author{Carsten Deibel $^1$}
\author{Vladimir Dyakonov $^{1,2}$}
\email[Electronic mail: ]{dyakonov@physik.uni-wuerzburg.de}
\affiliation{$^1$ Experimental Physics VI, Faculty of Physics and Astronomy, Julius-Maximilian-University of W\"urzburg, Am 
Hubland, 97074 W\"urzburg, Germany}
\affiliation{$^2$ Bavarian Center of Applied Energy Research e.V. (ZAE Bayern), Am Hubland, 97074 W\"urzburg, Germany}

\begin{abstract}
The trap states in three fullerene derivatives, namely PC$_{61}$BM ([6,6]-phenyl C61 butyric acid methyl ester), bisPC$_{61}$BM (bis[6,6]-phenyl C61 butyric acid methyl ester) and PC$_{71}$BM ([6,6]-phenyl C71 butyric acid methyl ester), are investigated by thermally stimulated current measurements (TSC). Thereby, the lower limit of the trap densities for all studied methanofullerenes exhibits values in the order of $10^{22}$~m$^{-3}$ with the highest trap density in bisPC$_{61}$BM and the lowest in PC$_{61}$BM. Fractional TSC measurements on PC$_{61}$BM reveal a broad trap distribution instead of  discrete trap levels with activation energies ranging from 15~meV to 270~meV and the maximum at about 75~meV. The activation energies of the most prominent traps in the other two fullerene derivatives are significantly higher, being at 96~meV and 223~meV for PC$_{71}$BM and 184~meV for bisPC$_{61}$BM, respectively. The influence of these findings on the performance of organic solar cells is discussed.\newline
 \textit{This is the pre-peer reviewed version of the following article:} J.SCHAFFERHANS, C. DEIBEL, AND V. DYAKONOV. ELECTRONIC TRAP STATES IN METHANOFULLERENES. ADVANCED ENERGY MATERIALS, 1:655, 2011 \textit{which has been published in final form at}\newline
[{\bf \href{http://dx.doi.org/10.1002/aenm.201100175}{Advanced Energy Materials 1, 655 (2011)}}].

\end{abstract}

\maketitle

\section{Introduction}
Methanofullerenes are the most commonly used electron acceptors in organic bulk heterojunction solar cells~\cite{deibel2010rev}. The first use of fullerene and its derivatives in organic photovoltaics (OPV) was introduced in~\cite{yu1995, sariciftci1992}. Since that, the methanofullerenes became the main type of acceptors in OPV. Also in the current record organic solar cell with efficiency of 8.3~\%~\cite{greenv37} a methanofullerene is used as acceptor. The advantages of methanofullerenes are that they can be easily processed from solution, possess high electron affinities and form segregated phases in blends with common donor polymers. Furthermore the used methanofullerenes yield good electron mobilities. For example, space--charge limited current mobilities for electrons of 2$\times$10$^{-3}$~cm$^2$/Vs in PC$_{61}$BM diode structures are reported~\cite{mihailetchi2003}. 

To further enhance the power conversion efficiencies of organic solar cells, a recent approach is to use multiple adduct fullerene derivatives~\cite{lenes2008, he2010}, such as  bisPC$_{61}$BM. Multiple side chains on the fullerene cage lead to an increase of the lowest unoccupied molecular orbital resulting in a raise of the open circuit voltage. 

Despite the importance of methanofullerenes, little attention was paid to thin films of the pure materials except the mobility measurements in n-type~\cite{singh2005, woebkenberg2008} or ambipolar~\cite{anthopoulos2004, anthopoulos2005} transistors and methanofullerene diodes~\cite{mihailetchi2003, lenes2008}. 

Recently, quantum chemical and voltammetric studies have been performed for PC$_{61}$BM and its higher fullerene adducts ~\cite{frost2010}, revealing a variety of HOMO and LUMO energies for the adducts with multiple side chains, due to the different isomers they consist of.

Lenes et al.~\cite{lenes2009} investigated electron--only devices based on  fullerene derivatives and a variety of their bisadducts blended with poly(3-hexylthiophene) by current--voltage measurements.  The lower currents of the devices with bisadducts were attributed to shallow trapping. This conclusion was supported by device simulations introducing a distribution of traps, which enabled to fit the current--voltage curves. 

Although, trap states can have a significant influence on the performance of organic solar cells, as they can act as recombination centers, lower the mobility and disturb the internal field distribution, direct identification of the trap states in methanofullerene films has not been performed so far. 

In this paper we investigate the trap states in three commonly used fullerene derivatives, namely PC$_{61}$BM, PC$_{71}$BM and bisPC$_{61}$BM, by thermally stimulated current measurements.

\newpage
\section{Results and Discussion}
\subsection{Experimental Results}
\begin{figure}[H]
	\centering
	\includegraphics[width=8cm]{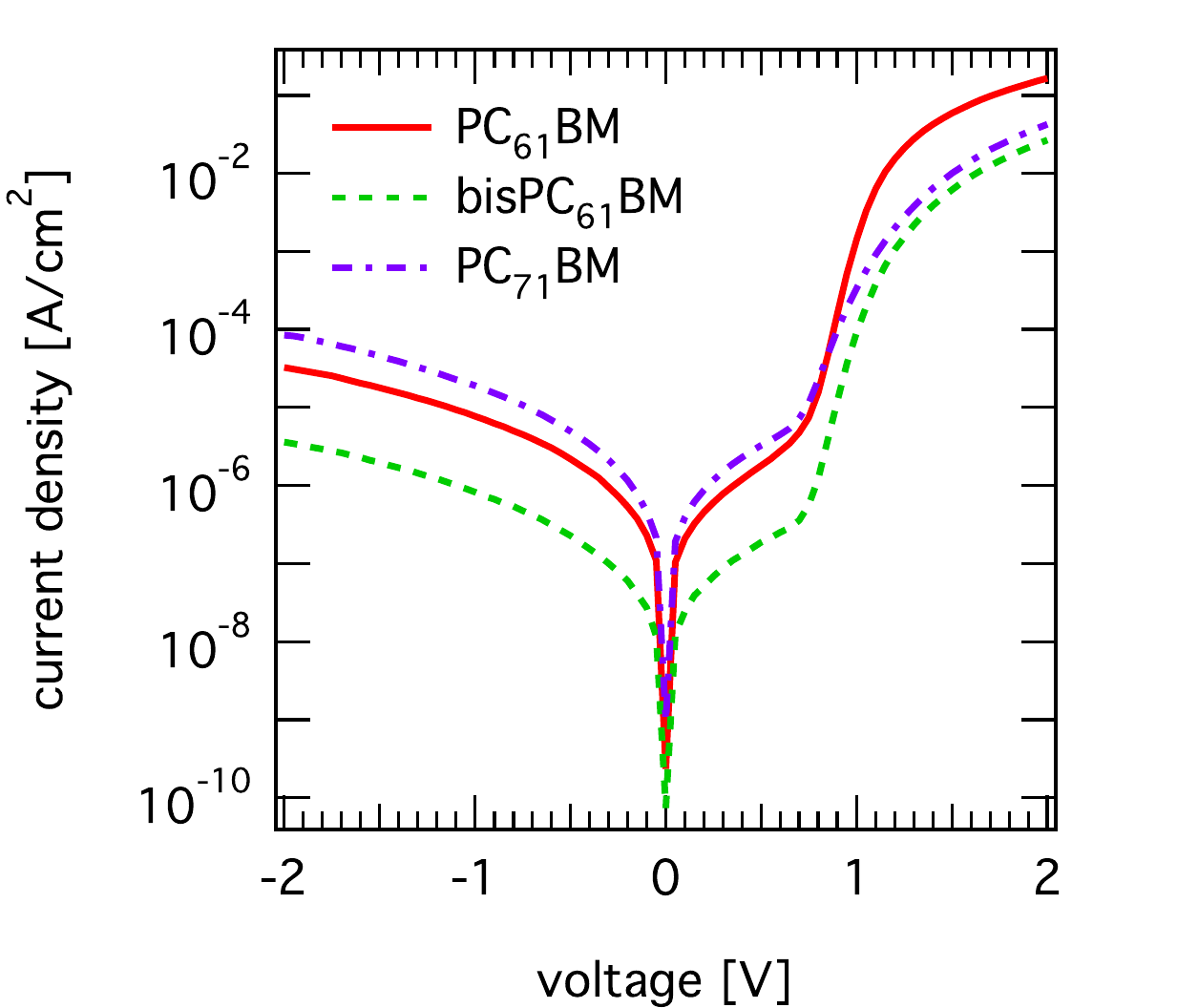}
	\caption{Dark IV--curves of PC$_{61}$BM, bisPC$_{61}$BM and PC$_{71}$BM diodes at room temperature. The diode structures were ITO/PEDOT:PSS/PCBM/LiF/Al and the thicknesses of all methanofullerene layers were about 200~nm.}
	\label{fig:IV-PCBMs}
\end{figure}

The current--voltage characteristics of the investigated samples are shown in Fig.~\ref{fig:IV-PCBMs}. 
The device currents in forward bias of bisPC$_{61}$BM and PC$_{71}$BM are lower compared to PC$_{61}$BM, with the lowest current for bisPC$_{61}$BM. 

\begin{figure}[htb]
	\centering
	\includegraphics[width=7.5cm]{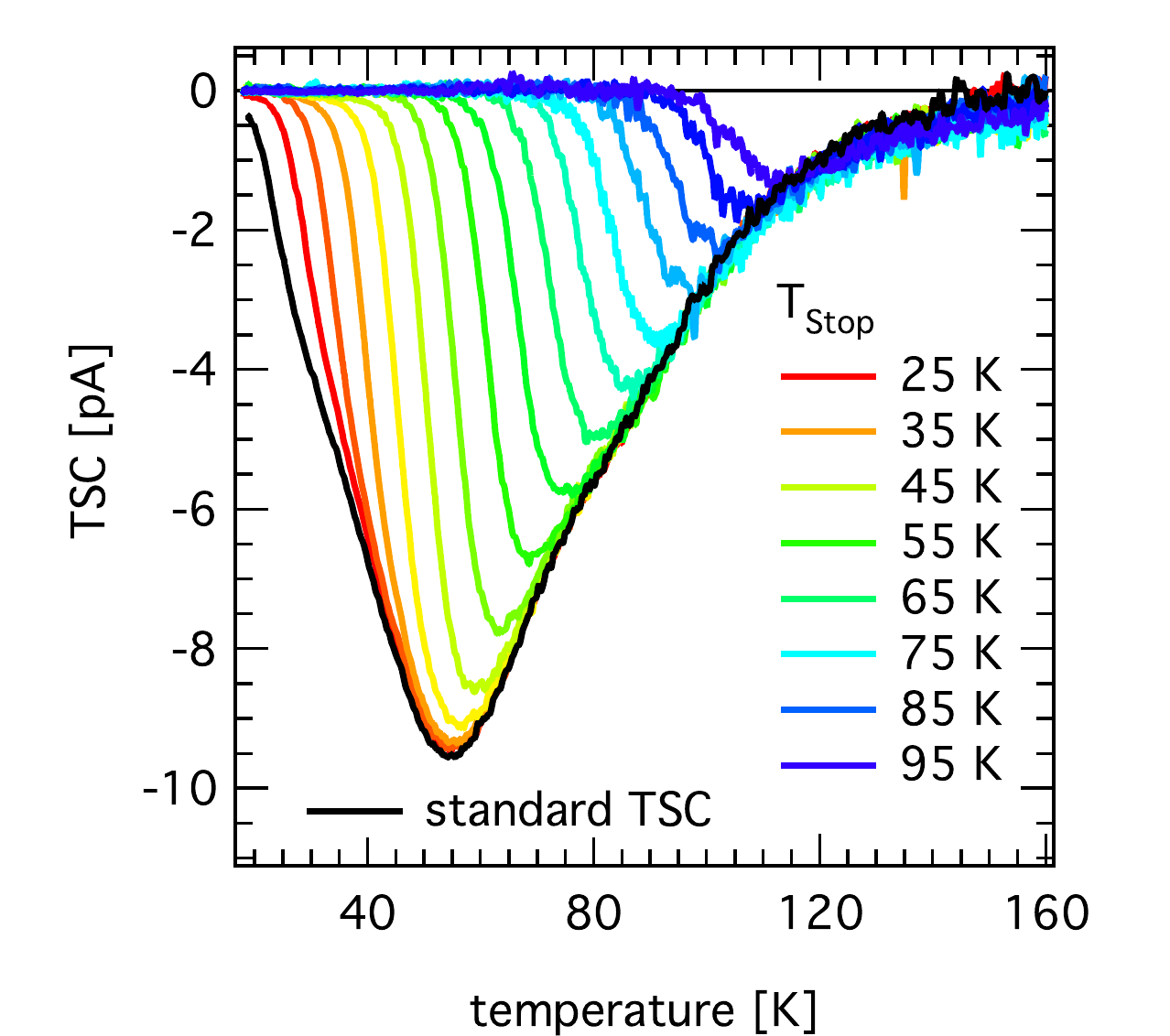}
	\caption{Standard TSC spectrum of PC$_{61}$BM (black line) as well as the fractional TSC spectrum by the $T_{Start}$--$T_{Stop}$ method for different $T_{Stop}$. }
	\label{fig:StartStop}
\end{figure}

The TSC spectrum of the PC$_{61}$BM sample is shown in Fig.~\ref{fig:StartStop} (black line). In this experiment, the current is measured upon warming up the device (see Experimental part). TSC behavior is observed between 18~K and 160~K with the maximum at about 55~K. No additional TSC peak can be detected at higher temperatures up to room temperature. Already at 18~K a small current of about 0.5~pA can be measured, indicating a release of trapped charge carriers. In addition to the maximum of the TSC spectrum at 55~K, small shoulders can be seen on the low temperature side (at about 25~K) as well as on the high temperature side (at about 90~K). The shape of the TSC spectrum indicates a trap distribution instead of discrete trap levels. A lower limit of the trap density n$_{t}$ can be obtained by integrating the TSC spectrum over the time, according to the inequality~\cite{kadashchuk2005}:

 \begin{equation}
	\centering
		\int_{peak} I_{TSC} dt \le e   n_{t}   V  ~~,
 	\label{eq:numtraps}
\end{equation}

with the elementary charge $e$ and the sample volume $V$. The inequality~(\ref{eq:numtraps}) is justified by the fact that  the estimated trap density is only a lower limit of the actual one. Incomplete trap filling, partial detrapping of charges during thermalization and recombination of the detrapped charge carriers of opposite signs are reasons therefor. According to inequality~(\ref{eq:numtraps}), the lower limit of the trap density of PC$_{61}$BM yields $n_t \ge$ 1.7$\times10^{22}$~m$^{-3}$.

\begin{figure}[htb]
	\centering
	\includegraphics[width=8cm]{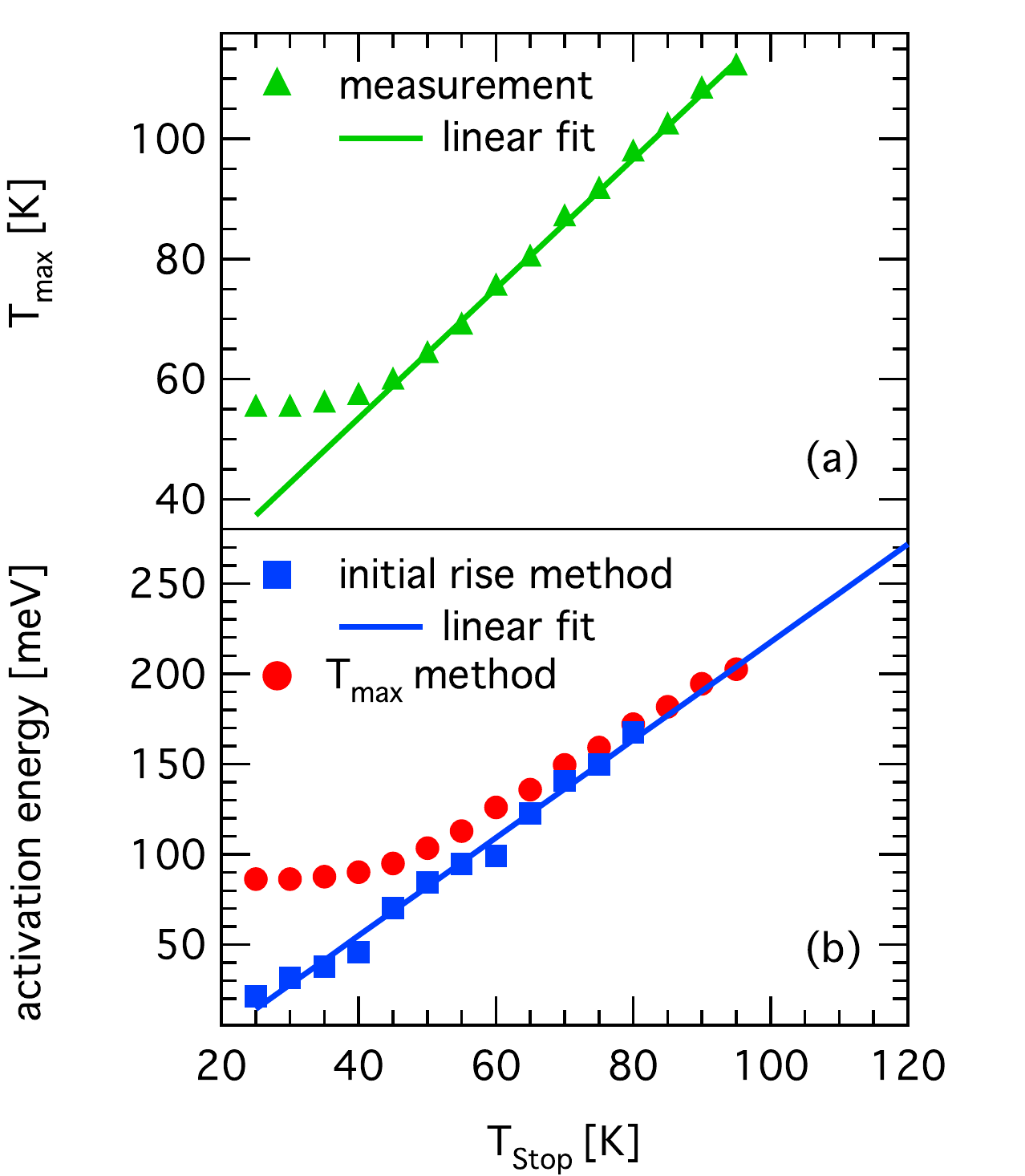}
	\caption{(a) Temperature of the TSC peak maximum of PC$_{61}$BM for different $T_{Stop}$ in the fractional TSC (triangles). The linear dependence of $T_{max}$ on $T_{Stop}$ (for $T_{Stop}$ $>$  40~K) of slope $\sim$~1 (line) indicates a quasi-continuous distribution of traps. (b) Activation energies of the trap states in PC$_{61}$BM according to the initial rise method (Equation~(\ref{eq:eact})) for the different $T_{Stop}$ (squares) and the linear fit of the activation energies (line). For comparison the activation energies obtained by the T$_{max}$ method (Equation~(\ref{eq:Tmax})) are also shown (circles).}
	\label{fig:TStop}
\end{figure}

To get further information about the trap distribution we applied fractional TSC measurements, the so called $T_{Start}$--$T_{Stop}$ method~\cite{schmechel2004, malm2002}. 
The illumination and thermalization of the sample was performed under the same conditions as for the standard TSC measurement, but instead of ramping the whole temperature range, the sample was only heated to a certain temperature $T_{Stop}$ (prerelease). Afterwards, the sample was cooled down to  $T_{Start}$ (18~K) and without further trap filling the whole TSC scan was performed (main run). 
This procedure was repeated for different  $T_{Stop}$, ranging from 25~K to 120~K  in steps of 5~K. Then, a standard TSC spectrum was recorded again, to ensure that no degradation of the sample during the measurements occurred. This TSC spectrum (not shown) is identical to the initial measurement. Therefore, a degradation of the sample during the  $T_{Start}$--$T_{Stop}$ cycles can be excluded. The main runs of the fractional TSC for different $T_{Stop}$ are shown in Fig.~\ref{fig:StartStop}. The shoulder at low temperatures becomes less pronounced for increasing $T_{Stop}$ due to emptying of the shallower traps during the prerelease and disappears for $T_{Stop}$ $>$ 35~K. The height and position of the main peak is unaffected  by the prerelease up to 40~K. For higher $T_{Stop}$ the peak maximum of the TSC decreases and shifts to higher temperatures. The maximum position of the TSC ($T_{max}$) is plotted vs $T_{Stop}$ in Fig.~\ref{fig:TStop}a (triangles), showing a linear dependence for $T_{Stop}$ $>$  40~K of slope $\sim$~1. 
A discrete trap level, in contrast, would result  in a constant maximum temperature, since only the number of trapped charges changes with $T_{Stop}$ but not the activation energy; therefore the TSC peak would remain at the same position. On the other hand, a series of trap levels, with well separated TSC peaks, would result in a staircase in the $T_{max}$--$T_{Stop}$ plot, with a flat region for each trap level. The closer are the trap levels and the more the TSC peaks overlap, the smoother becomes the staircase.  A linear dependence of $T_{max}$ on $T_{Stop}$, as seen in Fig.~\ref{fig:TStop}a,  is due to closely overlapping or a quasi-continuous distribution of TSC peaks~\cite{mckeever1980, mckeever} and reflects a quasi-continuous distribution of trap states in PC$_{61}$BM. With increasing $T_{Stop}$ the shallower traps get emptied and the influence of the deeper trap states on the TSC peak becomes stronger, resulting in a shift of the maximum to higher temperatures.

The activation energies $E_t$ can be estimated from the exponential initial rise~\cite{garlick1948} of the fractional TSC measurements for each  $T_{Stop}$ from:
\begin{equation}
	\centering
		I_{TSC} \propto \exp \left(-\frac{E_{t}}{k_{B} T} \right)~~,
	\label{eq:eact}
\end{equation}
where $I_{TSC}$ is the thermally stimulated current in the initial rise, $k_B$ the Boltzmann constant and $T$ the temperature. Hence, the activation energies for $T_{Stop}$  $\le$ 80~K  were determined from the Arrhenius plot, whereas the signal-to-noise ratio was too low for the cycles with $T_{Stop}$ above 80~K. The advantage of the initial rise method is that it can be applied without further knowledge of the kinetics of the TSC (i.e., if there is slow or fast retrapping of the charge carriers)~\cite{nicholas1964}. The obtained activation energies are shown in  Fig.~\ref{fig:TStop}b (squares). They show a good linear dependence on $T_{Stop}$, therefore, we linearly extrapolated the activation energies to estimate them in the range of $T_{Stop}$ $>$ 80~K.

\begin{figure}[htb]
	\centering
	\includegraphics[width=8cm]{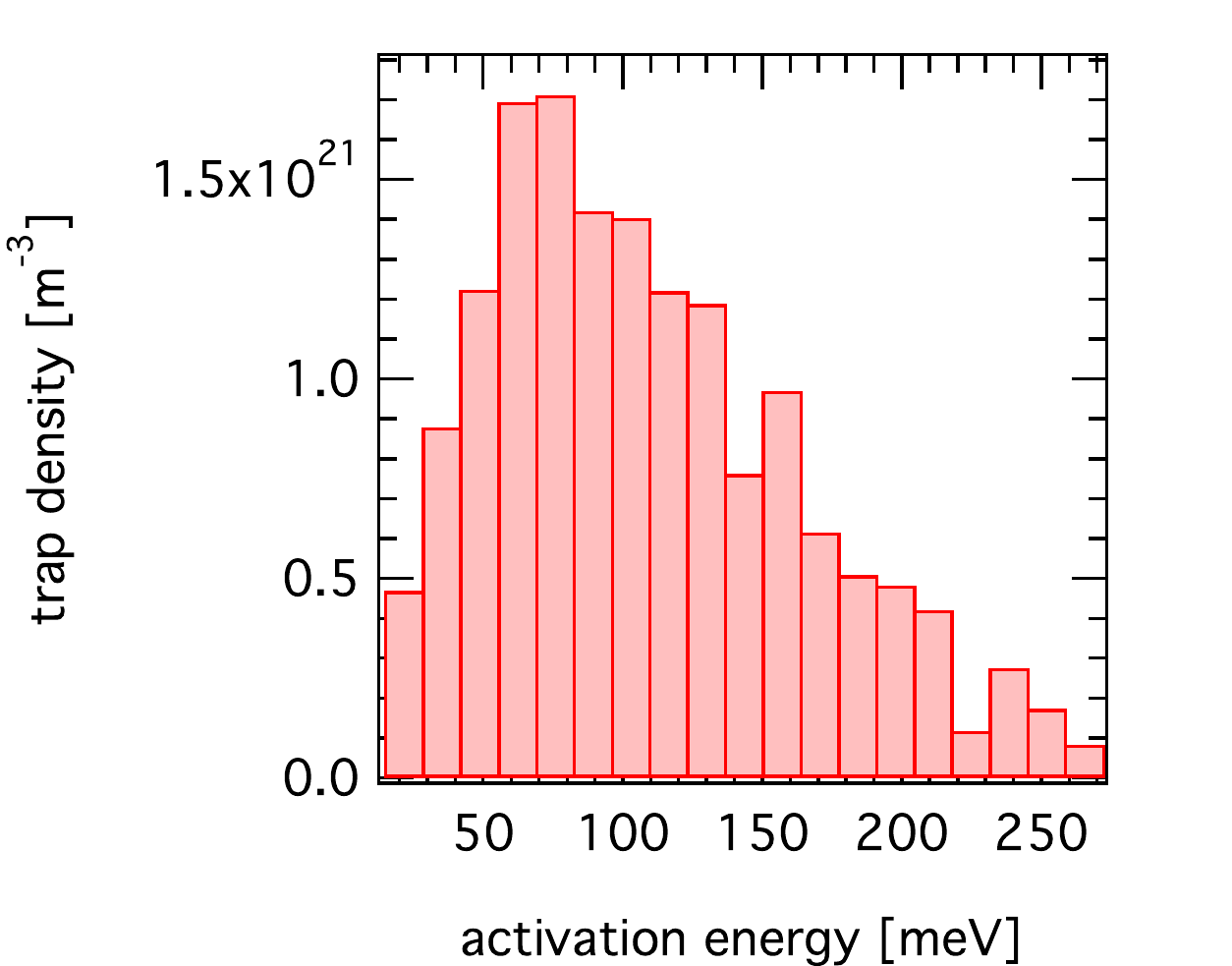}
	\caption{DOOS distribution of PC$_{61}$BM, as obtained by TSC $T_{Start}$--$T_{Stop}$ measurements.}
	\label{fig:trapdistribution}
\end{figure}

Applying inequality~(\ref{eq:numtraps}) to fractional TSC measurements, the trap densities for each $T_{Stop}$ interval can be estimated. These can be related to the activation energies extracted from the initial rise, as shown in Fig.~\ref{fig:trapdistribution}, yielding a reconstruction of the density of occupied states (DOOS)~\cite{schmechel2004}.  The histogram displays a broad trap distribution ranging from 15~meV to 270~meV with the maximum at about 75~meV.

The activation energy of the most prominent trap can also be approximately calculated using \cite{fang1990}:
 \begin{equation}
	\centering
		E_{t}=k_{B}  T_{max} \ln \left(\frac{T_{max}^4}{\beta} \right)  ~~.
 	\label{eq:Tmax}
\end{equation}
Here, $\beta$ is the heating rate used in the TSC measurements. This method (in the following called T$_{max}$ method) has also been used to estimate the activation energies of trap states in poly(3-hexylthiophene) (P3HT):PC$_{61}$BM solar cells~\cite{kawano2009} and C$_{60}$ field effect transistors~\cite{matsushima2007}. The advantage of this method is that no time consuming fractional measurements are needed. However, it  only gives an approximation of the dominating trap states, since otherwise distinct TSC maxima are needed. Nevertheless, to compare the trap states of different materials the information about the activation energies of the most prominent traps can be sufficient. Since the T$_{max}$ method is based on some assumptions and simplifications (e.g. retrapping of charge carriers is neglected)~\cite{nicholas1964, fang1990}, we first compare the activation energies of PC$_{61}$BM estimated by Equation~(\ref{eq:Tmax}) with the results of the initial rise method (Equation~(\ref{eq:eact})). As can be seen from Fig.~\ref{fig:TStop}b, the activation energy obtained with the T$_{max}$ method (circles) remains nearly constant for $T_{Stop}$ $<$ 40~K. This constant value of about 86~meV is caused by the fact that at low $T_{Stop}$  the temperature of the TSC peak maximum does not change, since the dominant traps are not affected by the prerelease, as already mentioned above. Hence, the activation energies obtained by the T$_{max}$ method are overestimated with respect to those from the initial rise method (squares), as the latter are dominated by the shallowest occupied traps. For $T_{Stop}$ $>$  40~K, the activation energy obtained from the T$_{max}$ method rises due to the emptying of the prominent traps and the enhanced contribution of  the remaining deeper traps. The difference between both methods diminishes. 

By applying Equation~(\ref{eq:Tmax}) to a standard (non-fractional) TSC scan one gets the activation energy of the most prominent traps. Therefore, the value of 86~meV (Fig.~\ref{fig:TStop}b) has to be compared to the energy distribution maximum shown in Fig.~\ref{fig:trapdistribution}. We may conclude, the T$_{max}$ method is a good approximation to gain the activation energy of the dominant trap states in methanofullerenes.

\begin{figure}[htb]
	\centering
	\includegraphics[width=8cm]{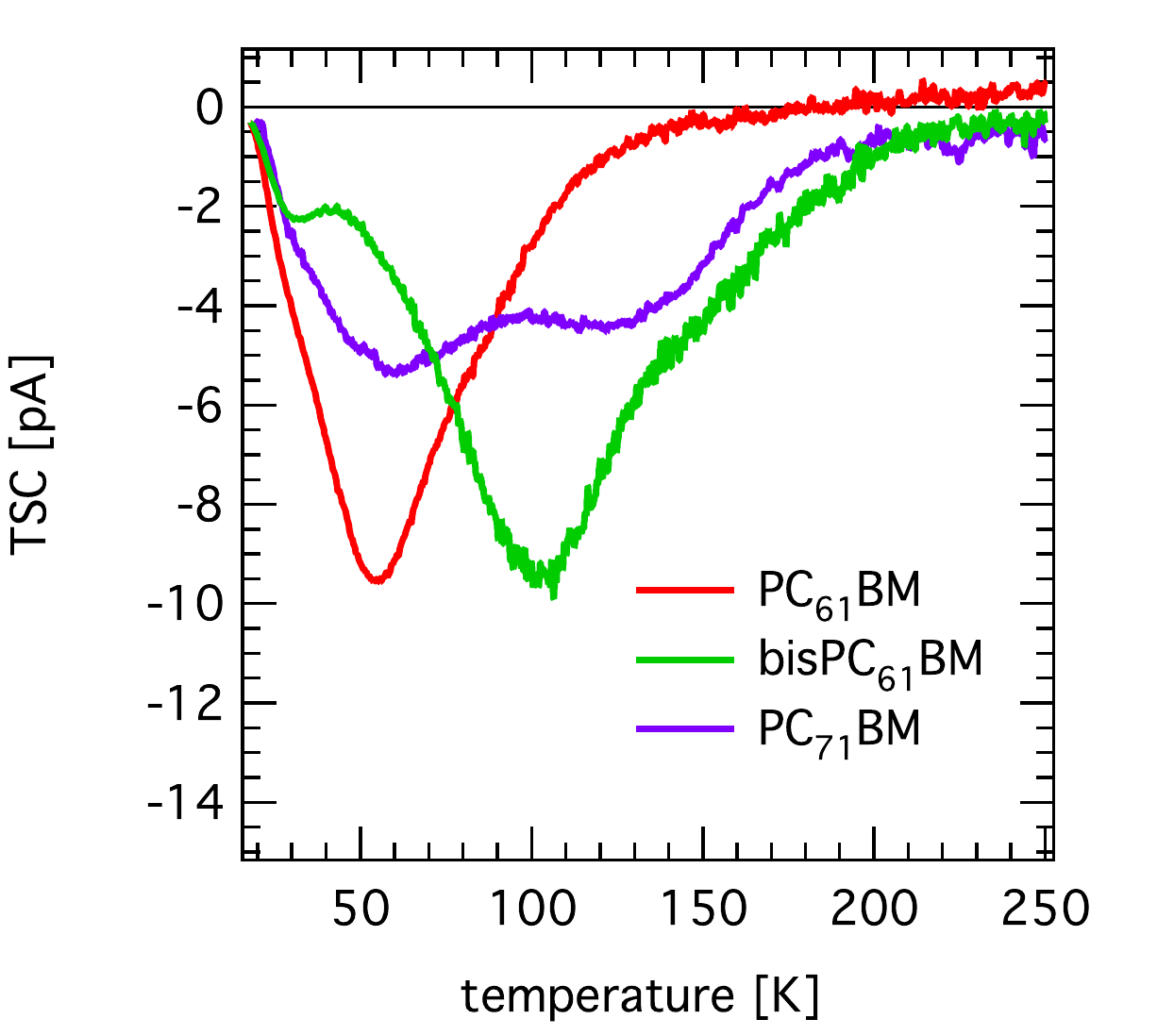}
	\caption{TSC spectra of PC$_{61}$BM, bisPC$_{61}$BM and PC$_{71}$BM }
	\label{fig:TSC-PCBMs}
\end{figure}

In addition to the investigations of PC$_{61}$BM we performed TSC measurements on bisPC$_{61}$BM and PC$_{71}$BM. The spectra are shown in Fig.~\ref{fig:TSC-PCBMs}, with PC$_{61}$BM for comparison. Both, bisPC$_{61}$BM and PC$_{71}$BM, yield an even broader TSC spectrum than PC$_{61}$BM, implying a broader trap distribution and higher activation energies. For bisPC$_{61}$BM two distinct maxima can be seen in the TSC spectrum, one at about 32~K and the dominant one at about 103~K, as well as a shoulder at about 160~K, indicating at least three different trap levels. PC$_{71}$BM yields two maxima with almost the same height at about 60~K and 120~K.

To give an estimation of the activation energies of the traps in bisPC$_{61}$BM and PC$_{71}$BM, we used the T$_{max}$ method according to Equation~(\ref{eq:Tmax}). The resulting activation energies of the distinct traps (the shoulder of bisPC$_{61}$BM is neglected), the corresponding temperatures of the TSC maxima, as well as the lower limits of the total trap densities are summarized in Table~\ref{tab:1}. 

\begin{table}[htb]
	\begin{center}
	\begin{ruledtabular}
	\begin{tabular}{llll}
	sample &  trap density [m$^{-3}$] & $T_{max}$ [K]& $E_{t}$ [meV] \\
	 \hline
	 PC$_{61}$BM& $\ge$ 1.7$\times10^{22}$ &54.9 & 86 \\
	bisPC$_{61}$BM& $\ge$ 2.3$\times10^{22}$ & 32.4 & 45 \\
	&  & 103 & 184 \\
	PC$_{71}$BM & $\ge$ 2.0$\times10^{22}$ & 60.1 & 96 \\
	 & & 121.4 & 223 \\
	\end{tabular}
	\end{ruledtabular}
	\end{center}
	\caption{Lower limit of the trap densities of PC$_{61}$BM, bisPC$_{61}$BM and PC$_{71}$BM, as well as the temperatures of the TSC maxima and the corresponding activation energies estimated by the T$_{max}$ method according to Equation~(\ref{eq:Tmax}). For bisPC$_{61}$BM and PC$_{71}$BM $E_{t}$ and $T_{max}$ values of both peaks are shown.}
	\label{tab:1}
\end{table}




\subsection{Discussion}
As summarized in Table~\ref{tab:1}, bisPC$_{61}$BM exhibits a trap density $n_{t}$~$\ge$ 2.3$\times10^{22}~$m$^{-3}$ which is about 35~\% higher than the value obtained for PC$_{61}$BM. Also the trap density of PC$_{71}$BM is higher than the one for PC$_{61}$BM. 
However, we emphasize that the estimated trap densities are only lower limits.

Furthermore, the three fullerene derivatives feature very different TSC spectra (Fig.~\ref{fig:TSC-PCBMs}).
BisPC$_{61}$BM yields a much broader TSC spectrum than PC$_{61}$BM. The small TSC peak of bisPC$_{61}$BM at about 32~K with corresponding activation energy of the traps of 45~meV is in good agreement with the low temperature shoulder of PC$_{61}$BM; it has therefore probably the same origin. However, the most prominent trap in bisPC$_{61}$BM exhibits an activation energy of about 184~meV which is considerably higher than the activation energy of the dominant trap in PC$_{61}$BM of about 86~meV. The shoulder on the high temperature side of the bisPC$_{61}$BM TSC peak indicates the presence of a significant amount of even deeper traps.
The high activation energies might originate from the mixture of isomers in bisPC$_{61}$BM. The lowest unoccupied molecular orbital energies for the different isomers range from -3.71~eV to -3.54~eV~\cite{frost2010}, introducing a higher energetic disorder---where the lowest lying states may act as traps. Furthermore, due to the additional side chains a close packing of the bucky\-balls is inhibited, which may result in a higher spatial disorder, also introducing additional traps.

PC$_{71}$BM  yields two distinct peaks in the TSC spectrum. The one at lower temperatures with an activation energy of about 95~meV  is quite similar to the main trap of  PC$_{61}$BM. The second one yields high trap activation energies of about 223~meV. Similar to bisPC$_{61}$BM, PC$_{71}$BM also consists of multiple isomers (one major and two minor isomers)~\cite{brabec2008book} which might be the origin of the deep traps. Furthermore, the PC$_{71}$BM batch has a lower purity grade ($>$99~\%) compared to PC$_{61}$BM ($\ge$99.5~\%) and therefore a higher impurity concentration which can lead to a higher trap density. Another explanation for the higher trap density and the deeper traps compared to PC$_{61}$BM might be a higher disorder in PC$_{71}$BM films, due to the spatial anisotropy of C$_{70}$.

The shapes of the TSC spectra (the shoulders in addition to the main peak) of the three investigated methanofullerenes indicate that the trap distribution of each of the fullerene derivatives consists of at least three peaks. This is affirmed by the fractional TSC measurements on PC$_{61}$BM yielding a quasi-continuous trap distribution between 15~meV and 270~meV. The TSC peaks of PC$_{71}$BM and bisPC$_{61}$BM are even broader compared to PC$_{61}$BM, which is in agreement with the reported higher energetic disorder of bisPC$_{61}$BM~\cite{lenes2009, frost2010}.

The higher trap densities and activation energies in bisPC$_{61}$BM and PC$_{71}$BM compared to PC$_{61}$BM are consistent with the observed lower currents in forward bias of these devices (Fig.~\ref{fig:IV-PCBMs}). In the case of bisPC$_{61}$BM this has also been observed before \cite{lenes2009} and was attributed to trapping. However direct measurements of the traps were missing so far.

Due to the high trap densities in the investigated methanofullerenes in the order of  $10^{22}$~m$^{-3}$, which is in the range of charge carrier densities of operating solar cells~\cite{shuttle2008b, rauh2011}, a strong influence on the solar cell performance can be expected.
Although, bisPC$_{61}$BM has successfully been used in P3HT:fullerene photovoltaic devices, a reduced photocurrent was observed compared to solar cells using PC$_{61}$BM as acceptor~\cite{lenes2008, lenes2009, faist2011}. Thereby, the reduced photocurrent is not due to a lower free polaron generation yield (as might be assumed because of the lower electron affinity of bisPC$_{61}$BM compared to PC$_{61}$BM). Even a higher charge photogeneration efficiency in the P3HT:bisPC$_{61}$BM blend was demonstrated by transient absorption spectroscopy~\cite{faist2011}. Instead, the loss in short circuit current was attributed to the significantly lower (about one order of magnitude) electron mobility in bisPC$_{61}$BM compared to PC$_{61}$BM~\cite{faist2011, lenes2008}. This increases the recombination probability during charge extraction and therefore reduces the photocurrent. The lower mobility of bisPC$_{61}$BM is in agreement with our measurements, showing a higher trap density and deeper trap states in bisPC$_{61}$BM  compared to PC$_{61}$BM. Furthermore, a slower decay of transient absorption intensity was observed in the blends with bisPC$_{61}$BM \cite{faist2011}. This was attributed to trapping and detrapping events resulting in a delay in the bimolecular recombination of charges. Our measurements reveal the presence of deeper trap states in bisPC$_{61}$BM, leading to a slower release of trapped charges, which is in agreement with the reported reduced recombination dynamics.

Protracted charge carrier decay dynamics, with a charge carrier density dependence larger than order of two, are often reported in organic solar cells \cite{shuttle2008, foertig2009} and are attributed to trapping of charges. Due to the high trap densities in the methanofullerenes, which are even higher than the reported trap density of  pristine P3HT \cite{schafferhans2008}, the recombination dynamics should be also significantly influenced by the acceptor, not only by the donor polymer. Because of the higher trap densities and deeper traps in bisPC$_{61}$BM and PC$_{71}$BM we expect even stronger dependences of the charge carrier decay on the charge carrier density  in blends with bisPC$_{61}$BM and PC$_{71}$BM than in those with PC$_{61}$BM.

\section{Conclusions}
We investigated the trap states in PC$_{61}$BM, PC$_{71}$BM and bisPC$_{61}$BM by thermally stimulated current measurements. The lower limit of the trap densities for each of the three methanofullerenes yields values in the order of $10^{22}$~m$^{-3}$ with the highest trap density for bisPC$_{61}$BM of $\ge$$2.3\times10^{22}$~m$^{-3}$. The activation energies of the most prominent traps in PC$_{71}$BM are 96~meV and 223~meV  and 184~meV for bisPC$_{61}$BM. Both reveal significantly deeper traps than PC$_{61}$BM, with an activation energy for the dominant trap of about 86~meV. These findings are consistent with the observed lower device currents in forward bias of the PC$_{71}$BM and bisPC$_{61}$BM diodes.  Additional fractional TSC measurements on PC$_{61}$BM revealed a broad trap distribution instead of  discrete trap levels with activation energies ranging from 15~meV to 270~meV.
 
 \section{Experimental Section}
The fullerene derivatives PC$_{61}$BM ([6,6]-phenyl C61 butyric acid methyl ester) (purity 99.5 \%), bisPC$_{61}$BM (bis[6,6]-phenyl C61 butyric acid methyl ester) (purity 99.5~\%) and PC$_{71}$BM ([6,6]-phenyl C71 butyric acid methyl ester) (purity 99~\%) in planar diode structures were used, prepared as follows. Structured indium tin oxide (ITO)/glass substrates were cleaned successively in soap water, acetone and isopropanol for at least 10~min in an ultrasonic bath. Afterwards, poly(3,4-ethylenedioxythiophene):(polystyrenesulfonate) (PEDOT:PSS) was spin coated on the substrates to serve as anode (thickness about 40~nm). After transferring the samples into a nitrogen filled glovebox a heating step of 130~$^\circ$C for 10~min was applied. The fullerene derivatives were spin coated from chloroform solution (PC$_{61}$BM: 30~mg/ml 600~rpm, bisPC$_{61}$BM and PC$_{71}$BM: 20~mg/ml 800~rpm). The layer thicknesses were about 200~nm as measured with a profilometer. LiF (1~nm)/Al (120~nm) top contacts were evaporated thermally (base pressure during evaporation $< 1\times10^{-6}$~mbar). The effective areas of the devices were about 9~mm$^2$. 
Fullerene derivatives were purchased from Solenne and used without further purification. PEDOT:PSS was purchased from H. C. Starck (CLEVIOS P VP Al 4083).

Initial current--voltage (IV) characteristics were measured in the nitrogen glovebox. 
Thermally stimulated current (TSC) measurements were performed in a closed cycle cryostat with Helium as contact gas. Via an integrated lock, the samples were transferred  from the glovebox to the cryostat, avoiding potential degradation of the samples due to air exposure. Trap filling was achieved by illumination of the samples at 18~K for five minutes using a 10~W high power white light emitting diode (Seoul). 
After a dwell time of five minutes the temperature was increased with a constant heating rate of 6.9~K/min up to 300~K. The TSC 
signals were detected by a Sub-Femtoamp Remote Source Meter (Keithley 6430) without applying an external field, implying that 
the detrapped charge carriers were extracted from the samples only due to the built-in voltage.  Further details of the TSC 
measurements are described elsewhere~\cite{schafferhans2008, schafferhans2010}.

\section*{Acknowledgements}
The current work is supported by the Bundesministerium f\"ur Bildung und Forschung in the framework of the OPV Stability Project 
(Contract No. 03SF0334F). J.S. thanks the Elitenetzwerk Bayern for funding. C.D. gratefully acknowledges the support of the Bavarian Academy of Sciences and Humanities. V.D.'s work at the ZAE Bayern is financed by the Bavarian Ministry of Economic Affairs, Infrastructure, Transport and Technology and by the Deutsche Forschungsgemeinschaft within the INST~93/557-1 project.

\end{document}